\documentclass[12pt, prd, showpacs]{revtex4}
\usepackage{amssymb}
\usepackage{amsmath}

\input{tcilatex}

\begin{document}

\title{Black hole with a scalar field as a particle accelerator}
\author{O. B. Zaslavskii}
\affiliation{Department of Physics and Technology, Kharkov V.N. Karazin National
University, 4 Svoboda Square, Kharkov 61022, Ukraine}
\affiliation{Institute of Mathematics and Mechanics, Kazan Federal University, 18
Kremlyovskaya St., Kazan 420008, Russia}
\email{zaslav@ukr.net }

\begin{abstract}
We consider stationary axially symmetric black holes with the background
scalar field and test particles that can interact with this field directly.
Then, particle collision near a black hole can lead to the unbounded energy $%
E_{c.m.}$ in the centre of mass frame (contrary to some recent claims in
literature). This happens always if one of particles is neutral whereas
another one has nonzero scalar charge. Kinematically, two cases occur here.
(i) A neutral particle approaches the horizon with the speed of light while
the velocity of the charged one remains separated from it (this is direct
analogue of the situation with collision of geodesic particles.). (ii) Both
particles approach the horizon with the speed almost equal to that of light
but with different rates. As a result, in both cases the relative velocity
also approaches the speed of light, so that $E_{c.m.}$ $\ $becomes
unbounded. We consider also a case when the metric coefficient $g_{\phi \phi
}\rightarrow 0$ near a black hole. Then, overlap between the geometric
factor and the presence of the scalar field opens additional scenarios in
which unbounded energy $E_{c.m.}$ is possible as well. We give a full list
of possible scenarios of high-energy collisions for the situations
considered.
\end{abstract}

\maketitle

\section{Introduction}

Several years ago, Ba\~{n}ados, Silk and West found that under certain
conditions, collision of particles near the Kerr black hole can give rise to
the unbounded energy $E_{c.m.}$ in the centre of mass frame \cite{ban}. This
is called the BSW effect after the names of its authors. It turned out that
for phenomenon of this kind, rotation is an essential ingredient, the effect
being universal \cite{prd}.\ Meanwhile, there are other versions and
analogues of this effect caused by the influence of the electric \cite{jl}
or magnetic \cite{fr} field. One more dynamic factor that potentially may
lead to unbounded $E_{c.m.}$ is the scalar field. In \cite{wei}, \cite{mao}
the possibility of unbounded $E_{c.m.}$ was examined for black holes with
the scalar (dilaton) field. However, the most part of corresponding results
can be considered as particular cases or slight modification of the general
scheme \cite{prd}, \cite{jl} since the scalar field acts as a source for
metric only, whereas particles themselves do not interact with this
background field.

Another interesting situation arises when collisions with unbounded $%
E_{c.m.} $ occur near naked singularities \cite{pjs}, \cite{fern}. It was
shown later \cite{sm} that the main reason giving rise to ultrahigh $%
E_{c.m.} $ in these papers, is related not to the scalar field (or any other
kind of a source) by itself but, rather, arises due to specific features of
the geometry near the horizon.

A possible dynamic role of the scalar field in the acceleration of particles
by black hole to ultrahigh $E_{c.m.}$ due to their interaction with the
background scalar field was examined, for the first time, in \cite{s1}, \cite%
{s2}. The situations considered in both these papers are quite different and
require separate attention. In \cite{s1}, collisions near static black holes
were studied. It was argued that the effect under discussion is absent in
this case. We show in the present paper that this is not so. In \cite{s1},
only collisions between two geodesic particles and two ones with the scalar
charge (charged, for brevity) were considered. The conclusion about
finiteness of $E_{c.m.}$ in both cases made in \cite{s1} is correct.
However, the most interesting case that does lead to the effect of ultrahigh
energies was overlooked. It consists of collision between a neutral particle
and a charged particle, as it will be seen below.

In \cite{s2}, similar scenarios were considered for a rotating metric with
the same conclusion. Meanwhile, in the end of \cite{s2}, a brief remark was
made that in the absence of interaction, unbounded $E_{c.m.}$ are possible
if both colliding particles are uncharged. This interesting properties has
no analogue in \cite{s1}. It was shown in \cite{sm} in the general setting
that the collision of two geodesic particles near the horizon indeed leads
to unbounded $E_{c.m.},$ provided the horizon is highly anisotropic (the
metric coefficient $g_{\phi \phi }\rightarrow 0$ where $\phi $ is the
azimuthal angle). The particular example considered in \cite{s2} (the
counterpart of the Kerr metric in the Brans-Dicke theory \cite{bdk1}, \cite%
{bdk2}, \cite{kim}) belongs just to this class of metrics and this explains
the result for the collision of two free particles. Thus two different
factors overlap in the example in Ref. \cite{s2} - anisotropy of horizons
and interaction between a scalar particle and the background scalar field.
As far as the role of the second factor is concerned, the main effect was
overlooked for rotating metrics as well as for static ones since collision
between the charged and neural particles (not considered in \cite{s2}) gives
rise to unbounded $E_{c.m.}$

The aim of the present work is to develop a general scheme describing high
energy collisions near scalar black holes. It applies to generic axially
symmetric rotating black holes. This includes also, as a particular case,
static black holes. In a model-independent approach, we show that collision
between a neutral and the charged particles near scalar black holes leads,
for an irregular background field, to unbounded $E_{c.m.}$. If,
additionally, $g_{\phi \phi }\rightarrow 0$, the overlap between geometric
and dynamic factors (due to the scalar field) allows also unbounded $%
E_{c.m.} $ for collision between two charged particles as well. We give a
full list of scenarios of high-energy collisions.

Another aim of the present work is to elucidate the general physical
mechanism and trace how interaction of a scalar particle with the background
scalar field leads to unbounded $E_{c.m.}$ We do not consider applications
to realistic astrophysics and discuss particular examples for the
illustrative purposes only.

The paper is organized as follows. In Sec. II, we list the general formulas
for the metric of a rotating axially symmetric black hole and equations of
motion for a particle interacting with a background scalar field in this
background. In Sec. III, we list general formulas for the energy in the
centre of mass frame of two colliding particles. \ In Sec. IV, we study the
near-horizon behavior of relevant quantities that enter the expression for $%
E_{c.m.}$. In Sec. V, we consider different scenarios of collisions an show
in a general form that (i) collision between two charged particles give
bounded $E_{c.m.}$, (ii) collisions between a neutral and a charged
particles lead to unbounded $E_{c.m.}$ In Sec. VI, we reveal the underlying
kinematic picture that explains the appearance of unbounded $E_{c.m.}$ In
Sec. VII, we discuss the behavior of the scalar field and a proper time for
near-horizon trajectories and relate them to the kinematic censorship that
forbids infinite energies. In Sec. VIII, the combined effect of dynamic \
and geometric factors is considered due to irregular scalar field and $%
g_{\phi \phi }\rightarrow 0$. In Sec. IX, we illustrate the obtained results
using the Bocharova, Bronnikov Melnikov and Bekenstein (BBMB) black hole 
\cite{bm}, \cite{73}, \cite{bek} as exact solution of field equations. We
show that particle collision in this background can indeed lead to unbounded 
$E_{c.m}\,$. In Sec. X, we exploit the exact solutions in the Brans-Dicke
theory as example and describe near-horizon behavior of relevant quantities
used in next section. In Sec. XI, we consider particle collisions in this
background and demonstrate explicitly that collision between a neutral and
the charged particles leads to unbounded $E_{c.m.}$ In Sec. XII, we give
summary of the results.

Throughout the paper, we put fundamental constants $G=c=1$.

\section{Rotating black holes}

Let us consider the metric%
\begin{equation}
ds^{2}=-N^{2}dt^{2}+g_{\phi }(d\phi -\Omega dt)^{2}+\frac{dr^{2}}{A}%
+g_{\theta }d\theta ^{2}\text{,}  \label{met}
\end{equation}%
where all coefficients do not depend on $t$ and $\phi $. We suppose that it
describes a black hole, so $N=0$ on the horizon. We assume the linear
interaction between a particle and the background scalar field $\psi $
described by the simplest action 
\begin{equation}
S=-\int (m+s\psi )d\tau ,  \label{S}
\end{equation}%
where $s$ is a coupling constant (the scalar charge of a particle), $m$ is
the mass, $\tau $ is the proper time. In what follows, we assume $s>0,$ so
that the combination $m+s\psi >0$ as well.

Equations of motion in this case read \cite{bek75}%
\begin{equation}
(m+s\psi )u_{;\beta }^{\alpha }u^{\beta }=-s[\psi ^{;\alpha }+u^{\alpha
}(\psi _{;\beta }u^{\beta })].  \label{eq}
\end{equation}

Here, $u^{\mu }=\frac{dx^{\mu }}{d\tau }$ is the four-velocity. Due to the
independence of the metric on $t$, there is the integral of motion that may
differ from the standard definition of energy due to interaction of a
particle with a scalar field:%
\begin{equation}
E=-(m+s\psi )u_{0}\text{.}  \label{e}
\end{equation}%
In a similar way, the angular momentum $L$ is also conserved,%
\begin{equation}
L=(m+s\psi )u_{\phi }\text{.}
\end{equation}%
For simplicity, we restrict ourselves by the motion in the equatorial plane $%
\theta =\frac{\pi }{2}$. Within this plane, one can always redefine the
radial coordinate to achieve $A=N^{2}$. Then, the equations of motion read
(dot denotes derivative with respect to $\tau $)%
\begin{equation}
m\dot{t}=\frac{\tilde{X}}{N^{2}}\text{,}  \label{tx}
\end{equation}%
\begin{equation}
m\dot{\phi}=\frac{\tilde{L}}{g_{\phi }}+\frac{\Omega \tilde{X}}{N^{2}}\text{.%
}
\end{equation}%
Here,%
\begin{equation}
\tilde{X}=\frac{Xm}{m+s\psi }\text{,}  \label{xt}
\end{equation}%
\begin{equation}
X=E-\Omega L\text{,}  \label{x}
\end{equation}%
\begin{equation}
\tilde{L}=\frac{Lm}{m+s\psi }.  \label{L}
\end{equation}

\begin{equation}
m\dot{r}=\sigma \tilde{Z}\text{,}  \label{mr}
\end{equation}%
where $\sigma =\pm 1$,%
\begin{equation}
\tilde{Z}=\sqrt{\tilde{X}^{2}-(m^{2}+\frac{\tilde{L}^{2}}{g_{\phi }})N^{2}}%
\text{.}  \label{zr}
\end{equation}

These equations can be formally obtained from equations of motion for
geodesic particles if one replaces $E$, $L$ and $X$ with their tilted
counterparts.

\section{Particle collisions: general formulas}

Now, let two particles 1 and 2 collide. One can define their energy in the
centre of mass \cite{ban} in the standard way:%
\begin{equation}
E_{c.m.}^{2}=-(m_{1}u_{1\mu }+m_{2}u_{2\mu })(m_{1}u_{1}^{\mu
}+m_{2}u_{2}^{\mu })=m_{1}^{2}+m_{2}^{2}+2m_{1}m_{2}\gamma \text{,}
\end{equation}%
where 
\begin{equation}
\gamma =-u_{1\mu }u^{2\mu }
\end{equation}%
is the Lorentz factor of relative motion. In what follows, we will consider
collision of two particle moving towards a black hole, so $\sigma
_{1}=\sigma _{2}=-1.$ Then, using the equations of motion listed above, one
obtains

\begin{equation}
m_{1}m_{2}\gamma =\frac{\tilde{X}_{1}\tilde{X}_{2}-\tilde{Z}_{1}\tilde{Z}_{2}%
}{N^{2}}-\frac{\tilde{L}_{1}\tilde{L}_{2}}{g_{\phi }}\text{.}  \label{ga}
\end{equation}

If, say, particle 1 is neutral, $\tilde{L}_{1}=L_{1}$, $\tilde{E}_{1}=E_{1}$%
, $\tilde{X}_{1}=X_{1}$.

\section{Near-horizon dynamics of charged particles}

We are interested in the near-horizon region since it is this region where a
small denominator in the first term in (\ref{ga}) may potentially lead to
unbounded $\gamma $. Dynamics of neutral (geodesic) particles in the context
of high energy collisions was studied in \cite{prd}. If a charged particle
participates in collision, the properties of $\gamma $ depend not only on
the metric but also on the behavior of the scalar field. If this field
remains bounded near the horizon, there is no qualitative difference between
the presence or absence of the scalar field, and we return to the known
situation to which the general analysis of \cite{prd} applies with minor
modifications. Therefore, we assume that near the horizon, the scalar field
diverges. For $N\rightarrow 0$, let the scalar field behave like%
\begin{equation}
\psi \approx cN^{-\beta }  \label{psb}
\end{equation}%
with 
\begin{equation}
\beta >0\text{,}  \label{be}
\end{equation}%
$c$ is some constant related to the scalar charge of the background
configuration. We assume that $c>0$, so $\psi >0$.

The fact that we deal with an infinite scalar field does not necessarily
mean that something pathological arises here. As is explained by J. D.
Bekenstein in \cite{bek75} (with the reference to private communication to
B. De Witt), "it is not associated with an infinite potential barrier for
test scalar charges; it does not cause the termination of any trajectories
of these test particles at finite proper time; and it is not connected with
unbounded tidal accelerations between neighboring trajectories". More
precisely, all this is also valid now for $\beta =1$. If $\beta <1$, the
acceleration diverges when the horizon is approached (see Sec. VII below).
Nonetheless, we include this case into consideration as well since in the
context of particle acceleration to unbounded energies all factors that
cause this effect (including the singular features in the metric or particle
dynamics) are of interest (see, e.g. \cite{pjs}, \cite{fern}, \cite{naked}).
It is also worth noting that divergent scalar field arises for some other
exact solutions that have a physical meaning of the counterparts of the Kerr
and Schwarzschild black holes in Brans-Dicke (see Sec. X and Sec. XI below).
In general, in scalar-tensor theories of gravity there is a freedom of
redefining the scalar field $\psi =\psi (\tilde{\psi})$ and using different
conformal frames according to $g_{\mu \nu }=\tilde{g}_{\mu \nu }e^{2\rho }$,
so what was finite in one conformal frame can in general be divergent in
another one and vice versa (in particular, on the black hole horizon). To
fix the class of frame to which our consideration applies, we assume that in
the frame where the asymptotic behavior (\ref{psb}) is valid, the
interaction between a particle and the scalar field is described by the
action (\ref{S}).

Now, near the horizon $\tilde{L}\rightarrow 0$ according to (\ref{L}). We
also assume that%
\begin{equation}
\lim_{N\rightarrow 0}g_{\phi }\neq 0\text{.}  \label{gf}
\end{equation}%
(The case when (\ref{gf}) is violated is considered in Sec. VIII and X, XI
below, see also \cite{sm}.) There are two possible cases to be discussed
separately (subscript "H" denotes quantities calculated on the horizon).
Below, we assume that a particle is charged.

1) $X_{H}\neq 0.$

If $\beta >1$, the negative term dominates (\ref{zr}), so that the condition 
$Z^{2}>0$ is violated. Therefore, instead, we assume that%
\begin{equation}
0<\beta \leq 1\text{.}  \label{bet}
\end{equation}

Let $\beta <1$. Then, near the horizon we have from (\ref{xt}), (\ref{psb}) 
\begin{equation}
\tilde{X}\approx \frac{X_{H}m}{cs}N^{\beta },  \label{xbe}
\end{equation}%
so $mN\ll \tilde{X}$. Taking also into account that $\frac{\tilde{L}^{2}}{%
g_{\phi }}\ll m^{2}$, we obtain 
\begin{equation}
\tilde{Z}\approx \tilde{X}-\frac{m^{2}N^{2}}{2\tilde{X}}\approx \tilde{X}-%
\frac{mcs}{2X_{H}}N^{2-\beta }\text{, }  \label{zn}
\end{equation}%
$\tilde{X}-\tilde{Z}=O(N^{2-\beta })$.

If $\beta =1$, 
\begin{equation}
\tilde{X}\approx \frac{X_{H}m}{cs}N,  \label{x1}
\end{equation}%
\begin{equation}
\tilde{Z}\approx Nm\sqrt{\frac{X_{H}^{2}}{c^{2}s^{2}}-1}\text{.}  \label{z1}
\end{equation}

2) $X_{H}=0$

We can use the near-horizon expansion for $X$. For extremal black holes \cite%
{dirty}, it reads 
\begin{equation}
X=B_{1}LN+O(N^{2})\text{,}  \label{xcr}
\end{equation}%
where $B_{1}$ is a constant and we took into account that $X_{H}=0$. By
substitution into (\ref{zr}), we formally obtain%
\begin{equation}
Z^{2}\approx m^{2}[\frac{L^{2}}{s^{2}c^{2}}N^{2+2\beta }(B_{1}^{2}-\frac{1}{%
g_{\phi H}})-N^{2}]\text{.}  \label{z0}
\end{equation}

The positivity of $Z^{2}$ is inconsistent with (\ref{be}) since the positive
term in (\ref{z0}) is much smaller than the negative one. This means that a
particle cannot reach the horizon, so we reject this case.

For nonextremal black holes \cite{dirty}, we would have%
\begin{equation}
X=C_{1}LN^{2}+O(N^{3})\text{,}  \label{ne}
\end{equation}%
where $C_{1}$ is a constant,%
\begin{equation}
Z^{2}\approx m^{2}(\frac{C_{1}^{2}L^{2}}{s^{2}c^{2}}N^{4+2\beta }-N^{2})-%
\frac{N^{2}\tilde{L}^{2}}{g_{\phi _{H}}}\text{,}
\end{equation}%
so it also would become negative and we arrive at the same conclusion as for
extremal black holes.

\section{When is $E_{c.m.}$ unbounded?}

Now, we apply the formalism under consideration looking for a possibility of
getting unbounded $E_{c.m.}$ in the horizon limit $N\rightarrow 0$. Let
particles have the coupling constants $s_{1}$ and $s_{2}$, respectively.

\subsection{$s_{1,2}\neq 0$}

It follows from the previous results that it is sufficient to consider the
case when $X_{H}\neq 0$ for both particles. We consider two situations
separately depending on the allowed value of $\beta $.

1) $\beta <1$.

Using (\ref{xbe}), (\ref{zn}) we obtain from (\ref{ga}) that%
\begin{equation}
\gamma \approx \frac{1}{2}[\frac{s_{2}}{s_{1}}\frac{\left( X_{H}\right) _{1}%
}{\left( X_{H}\right) _{2}}+\frac{s_{1}}{s_{2}}\frac{\left( X_{H}\right) _{2}%
}{\left( X_{H}\right) _{1}}]  \label{g1}
\end{equation}%
is finite.

2) $\beta =1$.

Then, (\ref{x1}), (\ref{z1}), (\ref{ga}) give us that%
\begin{equation}
\gamma \approx \frac{X_{1}X_{2}-\sqrt{X_{1}^{2}-c^{2}s_{1}^{2}}\sqrt{%
X_{2}^{2}-c^{2}s_{2}^{2}}}{c^{2}s_{1}s_{2}}  \label{gb1}
\end{equation}%
is finite again.

\subsection{$s_{1}=0$, $s_{2}=s\neq 0$}

In the same manner, we obtain that

1) $\beta <1$

\begin{equation}
\gamma \approx \frac{\left( X_{1}\right) _{H}cs}{2\left( X_{2}\right)
_{H}m_{1}}N^{-\beta }.  \label{g2}
\end{equation}

2) $\beta =1$ 
\begin{equation}
\gamma \approx \frac{\left( X_{1}\right) _{H}}{m_{1}Ncs}[\left( X_{2}\right)
_{H}-\sqrt{\left( X_{2}\right) _{H}^{2}-c^{2}s^{2}}]\text{.}  \label{gab}
\end{equation}

Thus both for $\beta <1$ and $\beta =1$ the Lorentz factor $\gamma $
diverges.

Thus, collisions between two charged particles or one charged and one
neutral particles lead to qualitatively different results. Is it possible to
pass from one case to another? From the formal viewpoint, this implies the
comparison of double limits $\lim_{N\rightarrow 0}\lim_{s_{1}\rightarrow
0}\gamma (N,s_{1},s_{2})$ and $\lim_{s_{1}\rightarrow 0}\lim_{N\rightarrow
0}\gamma (N,s_{1},s_{2})$. (For a moment, we showed explicitly the
dependence of the gamma-factor on relevant quantities.) We see from (\ref{g1}%
), (\ref{gb1}) and (\ref{g2}), (\ref{gab}) that, indeed, in both cases we
obtain divergent $E_{c.m.}$ To some extent, this resembles the situation
with double limits for the standard BSW effect - see eqs. (11), (15) in \cite%
{prd}.

It is seen from (\ref{mr}), (\ref{bet}), (\ref{zn}) and (\ref{z1}) that%
\begin{equation}
u^{r}=\dot{r}\rightarrow 0  \label{ur}
\end{equation}%
for any charged particle when $N\rightarrow 0$. This generalizes the
observation made in \cite{s2} for the Kerr-like solution in the Brans-Dicke
theory. Formally, there is also the combination $s_{1}=0=s_{2}$ but in this
case we return to the standard BSW context \cite{ban}, \cite{prd}.

\subsection{Collisions with participation of critical neutral particle}

Let us call a particle usual if $X_{H}\neq 0$ and critical if 
\begin{equation}
X_{H}=0.  \label{cr}
\end{equation}%
Up to now, we discussed collisions in which both particles are usual, so $%
\left( X_{1}\right) _{H}\neq 0$, $\left( X_{2}\right) _{H}\neq 0$. If \ a
charged particle is critical, it cannot approach the horizon at all, see eq.
(\ref{z0}) and subsequent discussion. However, this is not excluded for a
neutral particle. It is worth reminding that it is collision between a
critical and a charged particles gives rise to the standard BSW effect \cite%
{ban}, \cite{prd}. Therefore, it makes sense to examine also collision
between a charged usual one 1 and a neutral critical particle 2 ($\left(
X_{2}\right) _{H}=0$). Then, for particle 2 eq. (\ref{xcr}) is valid near
the horizon. By substitution into eq. (\ref{zr}), where $\tilde{X}=X$ and $%
\tilde{L}=L$, one has near the horizon%
\begin{equation}
Z_{2}\approx N\sqrt{(B_{1}^{2}-\frac{1}{g_{\phi }})L_{2}^{2}-m_{2}^{2}}\text{%
.}
\end{equation}

It is seen from that the case under discussion is reasonable for $\left(
g_{\phi }\right) _{H}\neq 0$ only. For $\beta <1$, using previous formulas (%
\ref{xbe}) and (\ref{zn}) for particle 1, one obtains after simple
transformations that%
\begin{equation}
m_{1}m_{2}\gamma \approx \left( X_{1}\right) _{H}\left( B_{1}L_{2}-\sqrt{%
(B_{1}^{2}-\frac{1}{g_{\phi }})L_{2}^{2}-m_{2}^{2}}\right) N^{\beta -1}\text{%
.}  \label{g12}
\end{equation}%
Taking into account condition (\ref{bet}) we see that the factor $\gamma $
diverges although somewhat slower that for the standard BSW effect due to an
additional factor $N^{\beta }$.

If $\beta =1$, eq. (\ref{x1}), (\ref{z1}) should be used for particle 1.
Then, it is seen that $\gamma $ remains finite.

For nonextremal black holes, $X_{1}=O(N^{2})$ (\ref{ne}) and according to (%
\ref{zr}), the condition $Z^{2}>0$ is violated. Therefore, such a neutral
particle cannot reach the horizon.

\section{Kinematic underlying reason for unbounded $E_{c.m.}$}

We consider a collision between a neutral (geodesic) particle 1 and the
charged one 2. For a geodesic particle moving in the background (\ref{met}),
there is the relation \cite{k}%
\begin{equation}
X=\frac{mN}{\sqrt{1-V^{2}}}\text{.}
\end{equation}%
Here, $V$ is the velocity measured by the local zero angular momentum
observer \cite{72}. Then, in the vicinity of the horizon, we have for
neutral particle 1:%
\begin{equation}
V_{1}^{2}\approx 1-\frac{mN^{2}}{\left( X_{1}\right) _{H}}\text{,}
\label{v1}
\end{equation}%
so $V_{1}\rightarrow 1$ in the limit $N\rightarrow 0$.

The tetrad components of velocity%
\begin{equation}
V_{1}^{(1)}=\sqrt{1-\frac{N^{2}}{X_{1}^{2}}(m_{1}^{2}+\frac{L_{1}^{2}}{%
g_{\phi }})}\text{,}  \label{11}
\end{equation}%
\begin{equation}
V_{1}^{(3)}=\frac{L_{1}N}{\sqrt{g_{\phi }}X_{1}}\text{,}  \label{13}
\end{equation}%
where $V^{(1)}$ is the component in the radial direction and $V^{(3)}$ is
that in the azimuthal direction (see \cite{k} for details). Near the
horizon, $V_{1}^{(3)}=O(N)\rightarrow 0$, 
\begin{equation}
V_{1}^{(1)}=1-O(N^{2}).  \label{1v}
\end{equation}%
A particle hits the horizon perpendicularly with the speed approaching that
of light.

According to the above explanations, the equations of motion for a charged
particle can be obtained from the geodesic ones by replacement of relevant
quantities with their tilted counterparts. As a result, we obtain:

\begin{equation}
\tilde{X}=\frac{Xm}{m+s\psi }=\frac{mN}{\sqrt{1-V^{2}}}\text{.}  \label{xn}
\end{equation}%
For particle 2,%
\begin{equation}
V_{2}^{(1)}=\sqrt{1-\frac{N^{2}}{\tilde{X}_{2}^{2}}(m_{2}^{2}+\frac{\tilde{L}%
_{2}^{2}}{g_{\phi }})}\text{.}  \label{1}
\end{equation}%
\begin{equation}
V_{2}^{(3)}=\frac{\tilde{L}_{2}N}{\sqrt{g_{\phi }}\tilde{X}_{2}}=\frac{L_{2}N%
}{\sqrt{g_{\phi }}X_{2}}\text{.}  \label{3}
\end{equation}%
Two subcases for the charged particle should be considered separately.

\subsection{$\protect\beta =1$}

The situation is completely similar to that in the standard case \cite{prd}, 
\cite{k}. Then, it follows from (\ref{x1}) and (\ref{1}), (\ref{3}) that
both components $V_{2}^{(3)}$ and $V_{2}^{(1)}$ are separated from zero. As
a result,the charged particle hits the horizon under some nonzero angle
relative to the normal direction, the absolute value $V_{2}<1$. Thus we have
collision between a rapid neutral particle and the slow charged one, so
explanation is completely similar to that for geodesics particles \cite{k}.
In doing so, the neutral particle in the scenario under discussion is a
counterpart of a usual one in the BSW effect, the charged particle
corresponds to the critical geodesic particle.

As a result, $\gamma $ diverges. According to (\ref{gab}), $\gamma
=O(N^{-1}) $.

\subsection{$\protect\beta <1$}

Now it is seen from (\ref{1}), (\ref{3}) that in the vicinity of the horizon 
$V_{2}^{(1)}\approx 1,V_{2}^{(3)}=O(N^{1-\beta })\rightarrow 0$, so $%
V_{2}^{(3)}\ll V_{2}^{(1)}$. Thus the charged particle, similarly to a
neutral one, hits the horizon perpendicularly. As far as the absolute value
of $V_{2}$ is concerned, we obtain from (\ref{xbe}) and (\ref{xn}) that 
\begin{equation}
V_{2}^{2}\approx 1-[\frac{sc}{\left( X_{2}\right) _{H}}]^{2}N^{2(1-\beta )}%
\text{.}  \label{v2}
\end{equation}%
Then, in the limit $N\rightarrow 0$ the velocity $V_{2}\rightarrow 1$.
Again, we see similarity with the case of a neutral particle. However, there
is also difference in that the charged particle approaches the horizon more
slowly than a neutral one because of the additional factor $N^{-2\beta }$ in
(\ref{v2}).

The behavior of the Lorentz gamma factor can be explained in terms of the
relative velocity $w$,%
\begin{equation}
\gamma =\frac{1}{\sqrt{1-w^{2}}}.
\end{equation}%
According to general formula (see. e.g., problem 1.3. in \cite{text}), the
relative velocity of particles $w$ obeys the relation%
\begin{equation}
w^{2}=1-\frac{(1-V_{1}^{2})(1-V_{2}^{2})}{(1-\vec{V}_{1}\vec{V}_{2})^{2}}%
\text{,}  \label{w}
\end{equation}%
where vectors and the scalar product are defined in the tangent space in
terms of tetrad components.

It follows from (\ref{11}), (\ref{13}), (\ref{1}), (\ref{3}) that%
\begin{equation}
1-\vec{V}_{1}\vec{V}_{2}=O(N^{2-2\beta })\text{.}
\end{equation}%
Then, using (\ref{v1}), (\ref{v2}), (\ref{w}), one finds that 
\begin{equation}
w^{2}=1-O(N^{2\beta })\text{.}
\end{equation}

Thus the relative velocity of two particles approaches the speed of light
and this is the reason why $E_{c.m.}$ grows unbounded.

For the case $\beta <1$ under discussion, both particles approach the
horizon almost with the speed of light but these two velocities do it with
essentially different rates. Therefore, in this case the kinematic mechanism
is different from that for the standard BSW effect \cite{k}.

\section{Proper time, acceleration and kinematic censorship}

The appearance of the unbounded energy leads to some subtleties. It is
clear, on physical ground, that in any process the relevant energy cannot be
infinite literally (the kinematic censorship), although it can be as large
as one likes. In the collision of two geodesic particles, one of such
particles is fine-tuned and this leads to an infinite proper time required
to reach the horizon \cite{gp-astro}, \cite{ted}, \cite{berti}, \cite{prd}.
Any actual collision occurs outside the horizon, $\ $so $\tau $ and $%
E_{c.m.} $ remain finite (although can be as large as one likes). What
happens in the present case?

According to (\ref{mr}), the proper time required for travelling between a
given point and the point of collision $r_{c}<r$ is equal to%
\begin{equation}
\tau =\int_{r_{c}}^{r}\frac{mdr}{\tilde{Z}}\text{.}  \label{tau}
\end{equation}

Let us consider the limit when $r_{c}\rightarrow r_{+}$, where $r_{+}$ is
the horizon radius. For a neutral particle it is finite since $Z_{H}\neq 0$.
For the charged particle, $\tilde{Z}=O(N^{\beta })$ according to (\ref{zn})
or (\ref{z1}). Let, for definiteness, the horizon be extremal, so 
\begin{equation}
N=O(r-r_{+}).  \label{nr}
\end{equation}
If $\beta =1$, it is seen from (\ref{tau}) that $\tau $ is infinite
similarly to the case when a particle is geodesic. However, for $\beta <1$
it is finite. How to reconcile infinite $E_{c.m.}$ with finite $\tau $?

The crucial point is the behavior of the acceleration $a_{\mu }$ of the
charged particle. For the action (\ref{S}), the acceleration is given by (%
\ref{eq}), hence%
\begin{equation}
a^{2}\equiv a_{\mu }a^{\mu }=\frac{s^{2}}{(m+s\psi )^{2}}h^{\mu \nu }\psi
_{,,\mu }\psi _{,\nu }\text{,}
\end{equation}%
\begin{equation}
h^{\mu \nu }=g^{\mu \nu }+u^{\mu }u^{\nu }\text{.}
\end{equation}%
For the extremal black hole (\ref{met}), evaluation of $a^{2}$ with (\ref{nr}%
) taken into account shows that $a$ is finite, provided $\beta =1$ in (\ref%
{psb}). However, if $\beta <1$,

\begin{equation}
a^{2}\sim N^{2(\beta -1)}
\end{equation}%
diverges. For nonextremal black holes, we have, by definition,%
\begin{equation}
N^{2}\sim r-r_{+}  \label{nn}
\end{equation}%
instead of (\ref{nr}). Then, divergences become even stronger, $a^{2}\sim
N^{2(\beta -2)}.$ Thus a particle experiences the action of infinite force
that should be considered as a singularity. Moreover, as the gradient of $a$
also diverges, any physical object of a finite size could not withstand such
infinite forces caused by the scalar field that would tear it. Therefore,
any actual collision should occur not exactly on the horizon but outside it,
so the kinematic censorship is preserved.

In the case $\beta =1$, the acceleration is finite, the proper time is
infinite, so the total configuration (geometry plus the scalar field)
exhibits no singular properties in spite of divergent scalar field (cf. \cite%
{bek75}). This is direct generalization of what happens in the case of the
BBMB black hole. If $\beta <1$, singular properties of the scalar field
reveal themselves in dynamics of a charged particle but the geometry itself
can be quite regular, neutral particles feel no singularity.

\section{Combined factors near horizon: divergent scalar field and vanishing 
$g_{\protect\phi }$}

As far as the role of the scalar field is concerned, we are faced with two
situations which are in sense complimentary to each other. Either (i) $%
\left( g_{\phi }\right) _{H}\neq 0$ and $\psi _{H}=\infty $ or (ii) $\left(
g_{\phi }\right) _{H}=0$ and $\psi _{H}<\infty $. Variant (i) is analyzed
above. Variant (ii) is the particular case of what was considered in \cite%
{sm}, where the material source (the scalar field or something else was
unimportant, provided $\left( g_{\phi }\right) _{H}=0$. Meanwhile, in \cite%
{pjs}, \cite{s2} one deals with the combination of both properties $\left(
g_{\phi }\right) _{H}=0$ and $\psi _{H}=\infty $. Therefore, for
completeness, we will consider such a case as well. To this end, we repeat
briefly the analysis of eq. (\ref{ga}) carried out in Sec. V. However, now
apart from (\ref{psb}) we must take into account also the condition $\left(
g_{\phi }\right) _{H}=0$. It is convenient to introduce the parameter%
\begin{equation}
b=\lim_{N\rightarrow 0}\frac{N}{\sqrt{g_{\phi }}}  \label{b}
\end{equation}%
similarly to what has been done in \cite{sm}.

We will assume that $\left( X_{H}\right) _{1}\neq 0$, $\left( X_{H}\right)
_{2}\neq 0$. If both particles are neutral, we return to the situation
already analyzed in Ref. \cite{sm}. It remains to study the following
combinations.

\subsection{Both particles are charged}

\subsubsection{$b\neq 0$}

Let in (\ref{psb}) $\beta =1$. It follows \ from (\ref{ga}) that in the
horizon limit $N\rightarrow 0$ (\ref{x1}) is valid and%
\begin{equation}
\tilde{Z}\approx \frac{Nm}{cs}\sqrt{X_{H}^{2}-b^{2}L^{2}},
\end{equation}%
where it is implied that $b\left\vert L\right\vert \leq X_{H}$ for each
particle. One can see that the Lorentz factor $\gamma $ remains finite.

Let $\,0<\beta <1$. Then, (\ref{xbe}) is still valid. One also obtains that
for each particle%
\begin{equation}
\tilde{Z}\approx N^{\beta }\frac{m}{cs}\sqrt{X_{H}^{2}-b^{2}L^{2}}\text{.}
\end{equation}%
As a result,%
\begin{equation}
\gamma \approx \frac{\left( X_{1}X_{2}-\sqrt{X_{1}^{2}-b^{2}L_{1}^{2}}\sqrt{%
X_{2}^{2}-b^{2}L_{2}^{2}}\right) _{H}-L_{1}L_{2}b^{2}}{N^{2-2\beta
}c^{2}s_{1}s_{2}}  \label{be<1}
\end{equation}%
diverges.

\subsubsection{$b=0$}

Let $g_{\phi }\rightarrow 0$ in the horizon limit in such a way that 
\begin{equation}
\frac{N^{2}}{g_{\phi }}\approx b_{1}^{2}N^{2\alpha }\text{, }0<\alpha \leq 1
\label{al}
\end{equation}%
near the horizon. Now, one can examine different possibilities.

If $1-\alpha \leq \beta \leq 1$, $\gamma $ turns out to be finite. If 
\begin{equation}
\beta <1-\alpha ,  \label{dba}
\end{equation}%
\begin{equation}
\gamma \approx \frac{b_{1}^{2}[(X_{2})_{H}L_{1}-(X_{1})_{H}L_{2}]^{2}}{%
2\left( X_{1}\right) _{H}\left( X_{2}\right) _{H}c^{2}s_{1}s_{2}}N^{2(\alpha
+\beta -1)}  \label{22}
\end{equation}%
diverges. We see that if $\left( g_{\phi }\right) _{H}=0$ a new possibility
opens for getting unbounded $\gamma $ that was absent for $\left( g_{\phi
}\right) _{H}\neq 0$ when collisions of two charged particles do not give
the effect under discussion.

The case of finite nonzero $\left( g_{\phi }\right) _{H}$ falls into this
scheme if we put $\alpha =1$ in (\ref{al}). Then, the necessary condition (%
\ref{dba}) for the unbounded $\gamma $ reduces formally to $\beta <0$.
Obviously, this cannot be realized since it is inconsistent with (\ref{be}).
Thus we confirm that collision of two charged \ particles near the horizon
with $\left( g_{\phi }\right) _{H}\neq 0$ cannot produce divergent $\gamma $
and hence $E_{c.m.}$

\subsection{Particle 1 is neutral, particle 2 is charged}

\subsection{$b\neq 0$}

Repeating calculations step by step, we arrive at the result 
\begin{equation}
\gamma \approx \frac{1}{scm_{1}}[X_{1}(X_{2}-\sqrt{X_{2}^{2}-b^{2}L_{2}^{2}}%
)-bL_{1}L_{2}]_{H}N^{\beta -2}\text{.}  \label{gb}
\end{equation}%
We see that $\gamma $ diverges according to (\ref{bet}).

\subsubsection{$b=0$}

We assume that the asymptotic form (\ref{al}) is valid. If condition (\ref%
{dba}) is fulfilled, one can infer that for particle 2%
\begin{equation}
\tilde{Z}\approx \frac{m_{2}N^{\beta }}{cs}[(X_{2})_{H}-\frac{b_{1}^{2}}{%
2(X_{2})_{H}}N^{2\alpha }L_{2}^{2}]\text{.}
\end{equation}%
For particle 1 we have from (\ref{zr}) with $\tilde{X}_{1}=X_{1}$ and $%
\tilde{L}_{1}=L_{1}$ and (\ref{al}) 
\begin{equation}
Z\approx X_{1}-N^{2\alpha }b_{1}^{2}L_{1}^{2}
\end{equation}%
Then, we obtain the following result from (\ref{ga}):%
\begin{equation}
\gamma \approx \frac{b_{1}^{2}[\left( X_{1}\right) _{H}L_{2}-\left(
X_{2}\right) _{H}L_{1}]^{2}}{2\left( X_{1}\right) _{H}(X_{2})_{H}m_{1}cs}%
N^{\beta +2(\alpha -1)}\text{.}  \label{12}
\end{equation}%
We see that $\gamma $ diverges.

If $\alpha +\beta >1$, the contribution of the term containing $g_{\phi }$
to $\tilde{Z}$ is negligible and we return to (\ref{g2}) for $\beta <1$ or (%
\ref{gab}) for $\beta =1$. In this sense, there is no difference between the
effect of unbounded $\gamma $ for nonzero or vanishing $\left( g_{\phi
}\right) _{H}$.

In the marginal case $\beta =1-\alpha $ we obtain that $\gamma =O(N^{-\beta
})$ that agrees with (\ref{12}) and (\ref{g2}), (\ref{gab}).

Thus collision between a charged particle and a neutral one does give the
unbounded $\gamma $ both for nonzero and zero $\left( g_{\phi }\right) _{H}$.

\subsection{Physical origin of unbounded Lorentz gamma-factor}

Both the special feature $\left( g_{\phi }\right) _{H}=0$ of the geometry
and the action of the scalar field can be considered as potential sources of
particles' acceleration. When they act together, one can ask: is it possible
to disentangle both factors and single out the main reason of the effect?
The answer depends strongly on a type of scenario. If both particles are
charged, characteristics of the geometry $b$, $b_{1}$, $\alpha $ enter the
expressions (\ref{be<1}), (\ref{22}) for $\gamma $ along with those of the
scalar field $c$, $s_{1}$,$s_{2}$. Therefore, it is impossible to
disentangle the roles of both factors that produce the combined effect.

Let us discuss now another scenario when a neutral particle collides with a
charged one. We see from (\ref{gb}) that $b\neq 0$ enters the expression for 
$\gamma $ as well as $c$ and $s$, so both factors are entangled as well as
in the previous case. If $b=0$, the situation is different. If (\ref{dba})
is valid, entanglement takes place also. However, if it is violated, the
final answer coincides with previous formulas (\ref{g2}), (\ref{gab}), so
the effect arises due to the scalar field only and has nothing to do with
the peculiarity of the geometry in question, according to which $\left(
g_{\phi }\right) _{H}=0$.

\section{Example with exact solutions: Bocharova, Bronnikov Melnikov and
Bekenstein (BBMB) black hole}

In this section, we illustrate general formalism using the example of
metrics that are exact solutions of field equations. As we saw, rotation did
not play an essential role in the effect under discussion. Therefore, to
simplify matter, we restrict ourselves with static metrics, so we put $%
\Omega =0$ in (\ref{met}). We choose here the solution describing the BBMB
black hole \cite{bm}, \cite{73}, \cite{bek}.\ Its metric can be written in
the form

\begin{equation}
ds^{2}=-dt^{2}f+\frac{dr^{2}}{f}+r^{2}d\omega ^{2}\text{,}  \label{bb}
\end{equation}%
where $d\omega ^{2}$ is the metric on unit 2-sphere, 
\begin{equation}
f=(1-\frac{M}{r})^{2}\text{.}
\end{equation}%
The scalar field%
\begin{equation}
\psi =\frac{q}{r\sqrt{f}}\text{,}
\end{equation}%
the quantity $q$ has the meaning of the scalar charge. In what follows, we
assume $q>0\,$, so $\psi >0$ as well. For the solution under discussion $q=%
\sqrt{\frac{3}{4\pi }}M.$

Although the solution under discussion is known to be unstable \cite{kir},
here we exploit it for the illustration of general formulas describing
particle collisions near a black hole. These formulas are insensitive to
whether or not a black hole is stable.

For simplicity, we will consider here pure radial motion within the plane $%
\theta =\frac{\pi }{2}$, so the angular momentum is equal to zero. Then, the
equations give us 
\begin{equation}
\dot{t}=\frac{\tilde{E}}{mf}\text{,}  \label{t}
\end{equation}%
\begin{equation}
\tilde{E}=\frac{Em}{m+s\psi }\text{.}  \label{E}
\end{equation}%
Using the normalization condition $u_{\mu }u^{\mu }=-1$, one obtains eq. (%
\ref{mr}) with%
\begin{equation}
\tilde{Z}=\sqrt{\tilde{E}^{2}-fm^{2}}\text{,}  \label{z}
\end{equation}%
It follows from (\ref{mr}) that%
\begin{equation}
f+\left( u^{r}\right) ^{2}=\frac{\tilde{E}^{2}}{m^{2}}\text{.}
\end{equation}

If $s=0$, there is no coupling between a particle and the background scalar
field. Then, $\tilde{E}=E$, $\tilde{Z}=Z=\sqrt{E^{2}-fm^{2}}$, and eqs. (\ref%
{t}), (\ref{mr}), (\ref{z}) turn into the geodesic equations.

We assume that particle 1 is neutral and particle 2 is charged. Using the
equations of motion, one finds from (\ref{ga}) that for small $f$%
\begin{equation}
\gamma \approx \frac{Y}{\sqrt{f}}\text{, }Y=\frac{E_{1}}{m_{1}}(\frac{E_{2}}{%
s}\sqrt{\frac{4\pi }{3}}-\sqrt{\frac{4\pi E_{2}^{2}}{3s^{2}}-1})\text{.}
\end{equation}%
It is implied that 
\begin{equation}
E_{2}\geq s\sqrt{\frac{3}{4\pi }},
\end{equation}%
otherwise the particle cannot approach the horizon. Thus $E_{c.m,}\sim
f^{-1/2}\rightarrow \infty $.

It is worth noting that one can use eq. (38) of \cite{s1} where collision
between particles with different scalar charges $f_{1}$ and $f_{2}$ (our $%
s_{1}$ and $s_{2})$ was considered. Then, one can take the limit $%
f_{1}\rightarrow 0$ while $f_{2}$ is kept fixed to confirm that $E_{c.m.}$
diverges in agreement with our general results.

In \cite{s1}, another example of exact spherically symmetric solution, the
so-called MTZ black hole \cite{mtz} was mentioned. For this solution, the
scalar field is regular on the horizon. Therefore, the described mechanism
does not work and $E_{c.m.}$ remains bounded not only for collision between
two charged or between two neutral particles \cite{s1} but also for
collision between a neutral and the charged one.

\section{Brans-Dicke analogue of Schwarzschild solution}

Here, we take advantage of the exact solution found within the Brans-Dicke
theory \cite{bdk1}, \cite{bdk2}, \cite{kim}. In general, it describes a
rotating black hole with a scalar field that represents the analogue of the
Kerr solution in the Brans-Dicke theory (the BDK solution). In the limit,
when rotation is absent, it turns to the counterpart of the Schwarzschild
solution (the so-called BDS solution). In a given context, this metric was
discussed in \cite{s2} with the conclusion that collision of charged
particles gives bounded $E_{c.m.}$ We would like to stress again that
collision between the charged and a neutral particles does indeed produce
unbounded $E_{c.m.}$.

Meanwhile, one should bear in mind the following subtlety. Now, as will be
seen below, the coefficient $g_{\phi }\rightarrow 0$ on the horizon. By
itself, this leads to the possibility of high energy collisions, even
without interaction with the background field ($s=0$) \cite{sm}. However,
now we are interested just in the role of this interaction, so one of
colliding particles is assumed to be charged. Thus in the case under
discussion, there is overlap between two completely different factors -
dynamic interaction with the background field and the properties of the
horizon geometry. Because of this overlap, the situation is a particular
case of what was considered in Sec. VIII.

By contrary to the standard BSW effect \cite{ban}, rotation is not essential
here, as is explained above and in \cite{sm}. Therefore, we restrict
ourselves by the nonrotating version of the BDK solution given by the BDS
one.

The metric can be written in the form (see eq. (13) of \cite{kim})%
\begin{equation}
ds^{2}=\Delta ^{-\frac{2}{2\omega +3}}\sin ^{-\frac{4}{2\omega +3}}\theta
\lbrack -dt^{2}(1-\frac{2M}{r})+r^{2}\sin ^{2}\theta d\phi ^{2}]+\Delta ^{%
\frac{2}{2\omega +3}}\sin ^{\frac{4}{2\omega +3}}\theta (\frac{dr^{2}}{1-%
\frac{2M}{r}}+d\theta ^{2})\text{,}
\end{equation}%
$\Delta =r(r-2M),$ the horizon is located at $r=r_{+}=2M$.%
\begin{equation}
\psi =\Delta ^{\frac{2}{2\omega +3}}\sin ^{\frac{4}{2\omega +3}}\theta .
\end{equation}

We restrict ourselves by motion within the plane $\theta =\frac{\pi }{2}$.
Then, equations of motion (\ref{tx}) - (\ref{zr}) read%
\begin{equation}
mu^{t}=\tilde{E}r^{2}\Delta ^{-\frac{2\omega +1}{2\omega +3}}\text{,}
\end{equation}%
\begin{equation}
mu^{\phi }=\frac{\tilde{L}}{r^{2}}\Delta ^{\frac{2}{2\omega +3}},
\end{equation}%
where $\tilde{E}$ and $\tilde{L}$ are given by eqs. (\ref{E}), (\ref{L}).
Eq. (\ref{zr}) takes the form%
\begin{equation}
\tilde{Z}=\sqrt{\tilde{E}^{2}-N^{2}(m^{2}+\frac{\tilde{L}^{2}}{r^{2}}\Delta
^{\frac{2}{2\omega +3}})}\text{,}  \label{zL}
\end{equation}%
where%
\begin{equation}
N^{2}=\frac{\Delta ^{\frac{2\omega +1}{2\omega +3}}}{r^{2}},  \label{n}
\end{equation}%
the scalar field is equal now to%
\begin{equation}
\psi =\Delta ^{\frac{2}{2\omega +3}}=r^{\frac{4}{2\omega +1}}N^{-\beta }%
\text{, }\beta =-\frac{4}{2\omega +1}\text{.}  \label{bed}
\end{equation}%
Now,%
\begin{equation}
\tilde{E}=\frac{Em}{m+s\Delta ^{\frac{2}{2\omega +3}}}\text{, }\tilde{L}=%
\frac{Lm}{m+s\Delta ^{\frac{2}{2\omega +3}}}\text{,}  \label{ed}
\end{equation}%
\begin{equation}
g_{\phi }=r^{2}\Delta ^{-\frac{2}{2\omega +3}}\text{.}  \label{gp}
\end{equation}%
Comparing this to (\ref{psb}), (\ref{bet}) and demanding that $N\rightarrow
0 $ when $\Delta \rightarrow 0$, we see that%
\begin{equation}
\omega \leq -\frac{5}{2}\text{.}  \label{5}
\end{equation}

Although the interval (\ref{5}) seems to be unrealistic from the
astrophysical point of view, we use it as a simple example to illustrate
general features of particle acceleration near scalar black holes.

It follows from (\ref{n}), (\ref{gp}) that eq. (\ref{al}) holds true with 
\begin{equation}
\alpha =\frac{2\omega +3}{2\omega +1}>0\text{.}  \label{alo}
\end{equation}

\subsection{Near-horizon behavior}

Near the horizon, $\Delta \rightarrow 0$. Then, for the charged particle 
\begin{equation}
\tilde{E}\approx \frac{mE}{s}\Delta ^{-\frac{2}{2\omega +3}}\text{, }\tilde{L%
}\approx \frac{mL}{s}\Delta ^{-\frac{2}{2\omega +3}}\text{.}
\end{equation}%
The term $\frac{\tilde{L}^{2}}{r^{2}}\Delta ^{\frac{2}{2\omega +3}}$ in (\ref%
{zL}) has the order $\Delta ^{-\frac{2}{2\omega +3}}$ and represents a small
correction to $m^{2}$ in (\ref{zL}), so it\ can be neglected.

If $\omega <-\frac{5}{2}$, we have 
\begin{equation}
\tilde{Z}\approx \tilde{E}-\frac{m^{2}}{2\tilde{E}}N^{2}\approx \frac{Em}{s}%
\Delta ^{-\frac{2}{2\omega +3}}-\frac{ms}{2Er_{+}^{2}}\Delta ,  \label{zc}
\end{equation}%
where the second term is a small correction to the first one. We see that
the term with $L$ does not affect the near-horizon expression for $Z$ in the
leading and sub-leading terms giving only unessential higher-order
corrections.

If $\omega =-\frac{5}{2}$ ($\beta =1$), 
\begin{equation}
\tilde{Z}\approx m\Delta \sqrt{\frac{E^{2}}{s^{2}}-\frac{1}{r_{+}^{2}}}\text{%
.}
\end{equation}

By contrast, for a neutral particle the near-horizon expansion depends on
the angular momentum more strongly. If $L=0$, 
\begin{equation}
Z\approx E-\frac{m^{2}}{2Er_{+}^{2}}\Delta ^{\frac{2\omega +1}{2\omega +3}}.
\label{nl}
\end{equation}

For a neutral particle with $L\neq 0$,%
\begin{equation}
Z\approx E-\frac{1}{2}\frac{L^{2}}{r_{+}^{4}E}\Delta \text{.}
\end{equation}

It turns out that the curvature invariants remain finite in the interval of
the Brans-Dicke parameter \cite{kim}%
\begin{equation}
-\frac{5}{2}\leq \omega <-\frac{3}{2}\text{.}  \label{int}
\end{equation}

According to (\ref{gp}), in the horizon limit $\Delta \rightarrow 0$ the
metric coefficient $g_{\phi }\rightarrow 0$ as well if $2\omega +3<0$ that
is compatible with (\ref{int}). However, the coefficient $g_{\theta \theta
}\rightarrow 0$, so that the horizon area $A=4\pi r_{+}^{2}$ remains finite.

As we are interested in the possibility of acceleration of particles by the
scalar field, eq. (\ref{5}) should be obeyed. In combination with (\ref{int}%
), it gives that either $\omega =-\frac{5}{2}$ (the horizon is regular) or $%
\omega <-\frac{5}{2}$ (then instead of a regular horizon we have a
singularity).

\section{Particle collisions in the BDS metric}

Now, we consider different possible situations.

\subsection{Both particles 1 and 2 are charged}

One can check that collisions between two charged particles gives the
bounded Lorentz factor $\gamma $ that agrees with \cite{s2}. It also agrees
with our previous general treatment since now 
\begin{equation}
\alpha +\beta -1=\frac{-2}{2\omega +1}>0  \label{ab1}
\end{equation}%
according to (\ref{bed}) and (\ref{alo}). Therefore, the necessary condition
(\ref{dba}) of unbounded $\gamma $ cannot be satisfied.

\subsection{Particle 1 is neutral, particle 2 is charged}

This is the most interesting case in our context not considered in \cite{s2}.

If $\omega =-\frac{5}{2}$, (\ref{ga}) gives us 
\begin{equation}
\gamma \approx \frac{r_{+}^{2}}{\Delta m_{1}}E_{1}(\frac{E_{2}}{s}-\sqrt{%
\frac{E^{2}}{s^{2}}-\frac{1}{r_{+}^{2}}})\text{,}  \label{gas}
\end{equation}%
so $\gamma $ and $E_{c.m.}$ diverge. Thus, in full analogy with the case of
the BBMB black hole (see above), the collision between a neutral and the
charged particles leads to unbounded $E_{c.m.}$

Let $\omega <-\frac{5}{2}$ (singular horizon). Then, using (\ref{ed}), (\ref%
{zc}) for a charged particle, we obtain that%
\begin{equation}
\gamma \approx \frac{E_{1}s}{2E_{2}m_{1}}\Delta ^{\frac{2}{2\omega +3}}
\label{gw}
\end{equation}%
diverges. It is worth noting that both for $\omega <-\frac{5}{2}$ and $%
\omega =\frac{5}{2}$ the values $L_{1}$ and $L_{2}$ do not enter the
asymptotic forms (\ref{gas}) and (\ref{gw})$.$

It follows from (\ref{ab1}) that, according to explanations given after eq. (%
\ref{12}), the results are insensitive to the fact that $\left( g_{\phi
}\right) _{H}=0$. Indeed, using expressions (\ref{bed}), (\ref{alo}) it is
easy to check that the results (\ref{gas}), (\ref{gw}) agree with general
formulas (\ref{gab}), (\ref{g2}) which were derived without account for $%
\left( g_{\phi }\right) _{H}=0$. In this sense, in the scenario under
consideration the unbounded $\gamma $ is achieved due to the scalar field
(that acts to particles 1 and 2 differently), whereas the effects of
curvature singularity are of the secondary importance. In particular, this
applies to the results of \cite{pjs}, \cite{s2}.

\subsection{Both particles 1 and 2 are neutral}

If at least one of angular momenta does not vanish, $\gamma =O(\Delta ^{%
\frac{2}{2\omega +3}})$ diverges with the same rate as in (\ref{gw}).
However, if $L_{1}=0=L_{2}$, it turns out that $\gamma $ is finite. This is
in agreement with a general case when in metric (\ref{met}) the coefficient $%
g_{\phi }\rightarrow 0$ on the horizon \cite{sm}.

In this context, it is the case $L_{1}=L_{2}=0$ for which the properties
under discussion become especially pronounced: the effect of unbounded $%
E_{c.m.}$ is absent both for collisions between two charged and two neutral
particles but reveals itself in collision between one charged and one
neutral particles.

\section{Discussion and conclusions}

In the present paper we gave a comprehensive analysis of the situations when
the effect of unbounded $E_{c.m.}$ in particle collisions can arise due to
interaction between the background scalar field and test scalar particles in
the vicinity of black holes. The summary of the results is given in Table 1
below, where we list only the scenarios capable to produce indefinitely
large $\gamma $ and $E_{c.m.}$. We also indicate which factor (the scalar
field or/and geometry) is relevant for a corresponding scenario. By
relevance of the geometric factor we imply that $\left( g_{\phi }\right)
_{H}=0$ that corresponds to $0<\alpha <1$ in the table according to (\ref{al}%
). If $\alpha =1$, $\left( g_{\phi }\right) _{H}\neq 0$. The index $\beta $
is responsible for the action of the scalar field. Thus the presence or
absence of $0<\beta \leq 1$ and $\alpha $ enables us to see which factor (or
both) is relevant in the effect of unbounded $\gamma $.

\begin{tabular}{|l|l|l|l|}
\hline
& Colliding particles & $g_{\phi }\sim N^{2(1-\alpha )}$, $0\leq \alpha \leq
1$ & Relevant factor(s) of two \\ \hline
1 & usual neutral and usual charged & $\left( g_{\phi }\right) _{H}\neq 0$, $%
\gamma \sim N^{-\beta }$ & scalar field \\ \hline
2 & usual neutral and usual charged & $\alpha +\beta >1$, $\gamma \sim
N^{-\beta }$ & scalar field \\ \hline
3 & usual neutral and usual charged & $\beta +\alpha \leq 1$, $\gamma \sim
N^{\beta +2(\alpha -1)}$ & scalar field and geometry \\ \hline
4 & usual charged and usual charged & $\alpha +\beta <1$, $\gamma \sim
N^{2(\alpha +\beta -1)}$  & scalar field and geometry \\ \hline
5 & critical neutral and usual charged & $\beta <1$, $\left( g_{\phi
}\right) _{H}\neq 0$, $\gamma \sim N^{\beta -1}$ & scalar field \\ \hline
6 & usual neutral and usual neutral & $\gamma \sim \frac{1}{g_{\phi }}\sim
N^{2(\alpha -1)}$ & geometry \\ \hline
7 & critical neutral and usual neutral & $\left( g_{\phi }\right) _{H}\neq 0$%
, $\gamma \sim N^{-1}$  & absent (standard BSW effect) \\ \hline
\end{tabular}

Table 1. Types of collisions that lead to unbounded $E_{c.m.}$

For completeness, we also put there in line 6 the results of our previous
work \cite{sm} and in line 7 the results inherent to the standard BSW effect 
\cite{ban}. They apply not only to the case of rotating black holes \cite%
{prd} but also to static electrically charged black holes (with minor
modifications of definitions of critical and usual particles - see \cite{jl}
for details). We see that inclusion of the mechanism under discussion due to
the scalar field gives rise to new scenarios of unbounded $\gamma $ (lines 1
- 5) that were absent in the standard case 7. Lines 1, 2 and 5 correspond to
the effect due to the scalar field in itself, line 6 represents the effect
due to the geometry, cases 3 and 4 give the combined effect of the scalar
field and the geometry when both factors cannot be disentangled from each
other. Main scenarios of collisions in the MDK metric considered in \cite{s2}
correspond to both usual charged particles with $\alpha +\beta >1$, so that
they do not lead to unbounded $\gamma $ and, therefore, do not fall into
this table. Meanwhile, the example typical of line 6 was mentioned in
passing in \cite{s2} after their eq. (24).

Thus the scalar field does act as a particle accelerator in that it leads to
new scenarios of high-energy collisions that were impossible without it.
Case 5 is somewhat special in that the scalar field accelerates the
particles but does it weaker than in the standard BSW effect 7. Some
suppression of collision energy takes place in case 5 but it does not cancel
the effect itself, provided $\alpha +\beta <1$. The latter condition is not
satisfied for the BDK solution and this is the reason of significant
suppression of collision energy because of which the effect of unbounded
energy $E_{c.m.}$ $\ $disappears \cite{s2}. In other words, the accelerator
due to the scalar field exists but it is not universal.

Cases 1 and 5 are the most interesting ones in that they are universal,
provided $\left( g_{\phi }\right) _{H}\neq 0$, so geometry of the horizon is
quite standard. Then, collision between a neutral and charged particles
always gives the effect of unbounded $\gamma $.

The results from Table 1 related to collisions between usual particles, are
valid for nonextremal and extremal black holes. This is not so for cases 5
and 7 since the approximate form of $X\,\ $for the critical particle in both
cases are different \cite{dirty} because of which a critical particle cannot
reach the nonextremal horizon at all \cite{prd}.

The research carried out in the present paper clearly revealed that a scalar
field can, in principle, be a particle accelerator. However, in contrast to,
say, the electromagnetic field \cite{jl}, it requires stronger conditions
such as divergence of the scalar field on the horizon.

An interesting issue for further research is whether and how the effect
under discussion reveals itself in cosmology where the scalar field can play
an essential role.

\begin{acknowledgments}
This work was funded by the subsidy allocated to Kazan Federal University
for the state assignment in the sphere of scientific activities.
\end{acknowledgments}

\end{document}